\documentclass[a4paper]{VisionStyle}
\usepackage{epsfig}
\newcommand{\lnls}{log\,N($>$S)-log\,S}
\begin{document}

\title{The XMM-Newton SSC Survey of the Galactic Plane}

\author{C.\,Motch\inst{1} \and X.\,Barcons\inst{2} \and F.\,Carrera\inst{2} \and P.\,Guillout\inst{1}
\and A.\,Hands\inst{3} \and B.J.M. Hassal\inst{4} \and O.\,Herent\inst{1} \and G.\,Lamer\inst{5} \and
A.\,Schwope\inst{5} \and G.\,Szokoly\inst{5} \and R.\,Warwick\inst{3} \and M.\,Watson\inst{3} \and N.
Webb\inst{6} \and  P.\,Wheatley\inst{3} } 

\institute{
  CNRS, Observatoire Astronomique de Strasbourg, 11 rue de l'Universit\'e, F-67000 Strasbourg, France
\and 
  Instituto de F\'isica de Cantabria (CSIC-UC), 39005 Santander, Spain
\and
  Department of Physics and Astronomy, University of Leicester, LE1 7RH, UK
\and
  Centre for Astrophysics, University of Central Lancashire, Preston, PR1, 2HE, UK
\and
  Astrophysikalisches Institut Potsdam, An der Sternwarte 16, D-14482 Potsdam, Germany 
\and 
  Centre d'Etude Spatiale des Rayonnements, 9 Av. Colonel Roche, F-31028, Toulouse, France
  }
\maketitle 

\begin{abstract} 

The XMM-Newton Survey Science Center is currently conducting an optical
identification programme of serendi\-pitous EPIC sources at low galactic latitudes.
The aim of this study is to quantify the various populations contributing to the
overall X-ray emission of the Galaxy and elaborate identification rules that can be
later applied to the bulk of the low galactic latitude EPIC detections. We report
here on preliminary results from an optical campaign performed in two
very low $b$ XMM-Newton fields and discuss the contributions of the various X-ray
populations. This paper is presented on behalf of the Survey Science Center and of
the AXIS collaboration.


\end{abstract}

\section{Introduction}

One of the most challenging responsibilities of the Survey Science Center (SSC) of the
XMM-Newton satellite is to characterise, in a proper fashion, the $\sim$ 50,000 new EPIC
sources discovered every year. In order to cope with this task, the SSC has designed a strategy
based on the concept of 'statistical identification'. Four samples of about 1000 EPIC sources
are the scope of optical observing campaigns aiming at a complete spectroscopic identification.
In parallel, multi-colour wide field imaging is carried out on a large number of XMM-Newton
fields. The correlations existing between the nature of the identified sources and their X-ray
and optical photometric characteristics will then be used to identify in a statistical manner
the rest of the XMM-Newton EPIC sources. At high galactic latitude, three spectroscopic samples
with limiting sensitivities of 10$^{-15}$, 10$^{-14}$  and 10$^{-13}$\,erg\,cm$^{-2}$\,s$^{-1}$
in the 0.5-4.5\,keV energy range are currently under investigation. First results from the
medium and bright sensitivity samples are discussed in \cite*{cmotch-WB1:bar02} and
\cite*{cmotch-WB1:dc02} and the overall role of the SSC and supporting optical observations are
described in \cite*{cmotch-WB1:wa01}. 

We report here on the status of the identification programme carried out on the fourth sample
located at low galactic latitude ($|b| \leq 20$\degr ), the XMM-Newton SSC
survey of the galactic plane (see \cite*{cmotch-WB1:mot01}). Many individual source types
contribute to the X-ray population of our Galaxy. These include late and early-type stars,
RS CVn's, cataclysmic variables (CVs), X-ray binaries, distant star forming regions and
supernovae remnants. Other more exotic X-ray emitters such as isolated neutron stars might
also show up. By separating these populations and eliminating the contamination due to
background AGN, we can hope to obtain a remarkable view of the accretion power, star
formation and end-points of stellar evolution throughout the Galaxy.

\section{X-ray and optical observations} 

In this paper we describe preliminary results from EPIC pn observations of two fields
belonging to the low galactic latitude sample. The first one, centered on the supernova remnant
\object{G21.5-0.9} was observed several times at different off-axis angles during the PV/Cal
phases. The source list discussed here was constructed from the two deepest observations
performed during revolutions 60 and 62. Usable EPIC pn + medium filter times are 19\,ksec and
23.5\,ksec respectively. The source search was performed on individual pointing data and later
merged into a single master list. The second field, Ridge~3, is located less than 2 deg away and
is slightly less deep with an EPIC exposure time of only $\sim$ 9.5 ksec. We list in Table
\ref{cmotch-WB1_tab:tab1} the main properties of the two target areas.

In order to be sensitive to the large variety of X-ray spectra exhibited by low latitude
sources both galactic and extragalactic, all detections made in the 0.5-2.0\,keV, 2.0-4.5\,keV,
4.5-7.5\,keV and 7.5-12\,keV bands were retained for optical identification. Source lists were
manually scree\-ned to reject false detections. In one field, G21.5-0.9, the positions derived
from the satellite attitude control system were further refined using sets of soft X-ray
sources identified with optically bright objects, presumably active coronae. Up to 4\arcsec\
offsets were applied, consistent with those found in high galactic latitude pointings
(\cite{cmotch-WB1:wa01}). 

Optical observations were made using telescope time allocated to the AXIS project which forms the
backbone of the SSC identification programme (\cite{cmotch-WB1:bar02a}). In addition, some wide-field
imaging data have been collected with the ESO-MPG 2.2m at La Silla. AXIS uses the largest telescopes of
the Observatorio del Roque de los Muchachos (Isaac Newton Telescope, Nordic Optical Telescope,
Telescopio Nazionale Galileo and William Herschel Telescope). Wide field u,g',r',i' and in some cases
also infra-red imaging were collected at the INT. Medium to low resolution spectroscopy was obtained
with the NOT, TNG and WHT telescopes using the ALFOSC, DOLORES, WYFFOS and ISIS instruments. 

\tabcolsep=4pt
\begin{table}[bht]
  \caption{Galactic XMM-Newton target fields}
  \label{cmotch-WB1_tab:tab1}
  \begin{center}
    \leavevmode
    \footnotesize
    \begin{tabular}[h]{lcccccc}
      \hline \\[-5pt]
      Field     & RA    & Dec     &  $l$ & $b$         &  No. of  & Area\\
                &   \multicolumn{2}{c}{(J2000)} & &    & sources  & (deg$^{2}$)\\[+5pt]
      \hline \\[-5pt]
      G21.5-0.9  & 18h33 & -10\degr\,34\arcmin & 21.5\degr & $-$0.9\degr & 70 &
      0.27\\
      Ridge~3  & 18h27 & -11\degr\,29\arcmin & 20.0\degr & +0.0\degr   &  21 &
      0.18\\
            \hline \\
      \end{tabular}
  \end{center}
  \vskip-1.5cm
\end{table}

\section{Optical identifications} 

The main criterion for identifying an X-ray source with an active corona is the presence of
Balmer and Ca II H\&K emission which are the commonly accepted signature of chromospheric
and hence coronal activity. Active M stars display easily detectable Balmer emission lines
even at relatively low signal to noise ratios. For solar type stars, Balmer emission may not
be detectable because of the underlying photospheric absorption and at the moderate
spectral resolution of 3 to 10\,\AA\ used in our optical identification work the Ca II H\&K
emission may have a too low contrast to be revealed in stars earlier than about G-K.
Therefore, part of the stellar identifications have to rely on the low probability to find
an optically bright object in the small EPIC error circle. In order to compute this
probability we used our R band imaging data and GSC 2.2 extractions to derive magnitudes
for the brightest optical objects in the 90\% confidence error circle as well as cumulative
stellar densities versus magnitudes. At the 95\% confidence level we estimate that any star
located in the 90\% error circle and brighter than R $\sim$ 17 and R $\sim$ 16.8 is a likely
counterpart for the \object{G21.5-0.9} and Ridge~3 fields respectively. The mean error on source
position was about 2\arcsec\ for these two fields. 

The current statistics of optical identifications is shown in Table
\ref{cmotch-WB1_tab:tab2}. The accreting source close to \object{G21.5-0.9} is the likely Be/X-ray
binary \object{SS 397} while the extragalactic source is the cluster \object{XMMU
J183225.4-103645} whose nature has been determined on the basis of XMM-Newton data alone
($z$ = 0.1242,  \cite{cmotch-WB1:ne01}). We estimate that we are optically complete down to
R $\sim$ 19 for emission line objects. 

\tabcolsep=6pt
\begin{table}[bht]
  \caption{Statistics of optical identifications}
  \label{cmotch-WB1_tab:tab2}
  \begin{center}
    \leavevmode
    \footnotesize
    \begin{tabular}[h]{lll}
      \hline \\[-5pt]
      Field     & G21.5-0.9        & Ridge~3  \\[+5pt]
      \hline \\[-5pt]
      Total    &  70        &   21   \\
      Stellar Coronae&  16 (23\%) &   11 (52\%)  \\
      Accreting      &   1        &    0      \\
      Extragalactic  &   1        &    0      \\
      Unidentified   &  52 (74\%) &    10 (48\%) \\
      \hline \\
      \end{tabular}
  \end{center}
  \vskip-0.5cm
\end{table}

Among the 27 active coronae are 8 Me stars. Most of them are fainter than R = 19, the
faintest one being R = 20.8. Fig. \ref{cmotch-WB1_fig:fig1} shows that the ratio of their
X-ray to H${\alpha}$ flux is very similar to those of Me stars detected in low galactic
latitude ROSAT surveys (\cite{cmotch-WB1:mo97}). So far, most of them fall in the same
spectral range (earlier than about M5) and the same X-ray luminosity range ({$\rm L_{\rm X}$
$\sim$ 10$^{29-30}$\,erg\,s$^{-1}$) as ROSAT detections. This indicates that the Me star
population detected by XMM-Newton is not much different from that seen in the ROSAT survey
and located roughly ten times farther away. The fraction of Me stars among XMM-Newton active
coronae ($\sim$ 30\%) may be slightly larger than that seen in the low latitude ROSAT survey
($\sim$ 19\%, \cite{cmotch-WB1:mo97}). Note, however, that the actual ratio may be somewhat
different since stellar identifications are certainly not yet complete.

\begin{figure}[ht]
  \begin{center}
  \epsfig{file=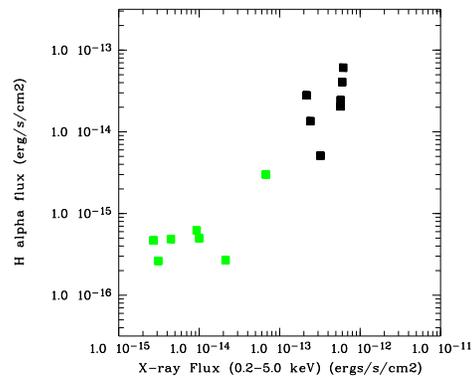, width=6.2cm, clip=true}
  \end{center}
  
\caption{H${\alpha}$ versus X-ray flux for Me stars. In green; Me stars identified in
the two XMM-Newton fields. In black; Me stars in the Cygnus area of the ROSAT galactic
plane survey. ROSAT to XMM-Newton flux conversion factor was derived assuming a thin thermal
spectrum with kT = 0.86\,keV and N$_{\rm H}$ = 5 10$^{\it 20}$\,cm$^{-2}$.}

\label{cmotch-WB1_fig:fig1}

\end{figure}

EPIC pn sources identified with active coronae display significantly softer HR2 than the
rest of the population (see Fig. \ref{cmotch-WB1_fig:fig2}). We use here the same
definition as in pipeline processing, i.e.; 
\medskip
\newline
\begin{math}
  {\rm HR2} = \frac{\rm C(2.0-4.5) - C(0.5-2.0)}{\rm C(2.0-4.5)+C(0.5-2.0)} 
\end{math}
\newline

\medskip 
where C(A-B) is the EPIC pn count rate in the A to B keV range. Our mean stellar HR2 is
 harder than that reported by \cite*{cmotch-WB1:dc02} (HR2 $\leq$ $-$0.8). Two
cooperating effects could account for harder coronal spectra in low $b$ samples. First, the
larger distances result in larger interstellar absorption toward the sources. Second,
stellar X-ray count models (\cite{cmotch-WB1:fa92}, \cite{cmotch-WB1:gu96}) predict that a
population of very young stars with small scale height or a population of very active
binaries (RS CVn's) may dominate stellar identifications in low latitude X-ray surveys. In
contrast, the high latitude population is expected to be mostly made of older and less X-ray
luminous stars. Therefore the overall correlation between X-ray luminosity and temperature
(see e.g. \cite{cmotch-WB1:sch97}) could also explain the harder stellar X-ray spectra seen
in our sample. 

The good correlation between HR2 and stellar nature gives us confidence
in the criteria applied and opens good prospects for setting up an automated
identification procedure for relatively bright active stars.

\begin{figure}[ht]
  \begin{center}
    \epsfig{file=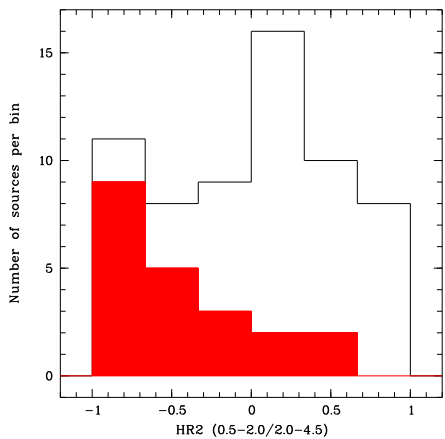, width=6.2cm, clip=true}
  \end{center}
  
\caption{HR2 distribution of all sources in the two fields. The red histogram represents the
distribution of X-ray sources identified with active coronae.}

\label{cmotch-WB1_fig:fig2}
\end{figure}

\section{The extragalactic source background}

The sensitivity of XMM-Newton above a few keV allows the extragalactic source background to
be detected throu\-ghout the whole Galaxy, even in the presence of relatively high
interstellar absorption. A good example of this 'contamination' is the discovery of an X-ray
bright cluster of galaxies (\object{XMMU J183225.4-103645}) in the field of
\object{G21.5-0.9} (\cite{cmotch-WB1:ne01}). Since the bulk of the photoelectric absorption
(N$_{\rm X}$ = 7.9 $\pm$ 0.5 10$^{22}$\,cm$^{-2}$) from the cluster is of interstellar origin
we can use it as a calibrator to estimate the range of integrated galactic N$_{\rm H}$ in the
two fields. The fact that the value of N$_{\rm HI}$ $\sim$ 2 10$^{22}$\,cm$^{-2}$
(\cite{cmotch-WB1:di90}) is much lower suggests that in these directions most of the
photoelectric absorption comes from the cold and dense phase of the ISM. The CO intensity at
the cluster location taken from the map of \cite*{cmotch-WB1:da01} (0.25 deg spatial
resolution) then yields a N$_{\rm H 2}$ to CO intensity  conversion factor of 2.7 $\pm$ 0.4
10$^{20}$\,cm$^{-2}$\,K$^{-1}$\,km$^{-1}$\,s. This value is slightly above that derived at $|b| \ \geq$
5\degr \ by \cite*{cmotch-WB1:da01} (1.8 $\pm$ 0.3 10$^{20}$\,cm$^{-2}$\,K$^{-1}$\,km$^{-1}$\,s) but
consistent with or below those listed in \cite*{cmotch-WB1:so86}. This 'local' calibration
and the CO velocity integrated map yield a total galactic N$_{\rm H}$ in the range of 5 to 9
10$^{22}$\,cm$^{-2}$ and 6.5 to 11.6 10$^{22}$\,cm$^{-2}$ for the \object{G21.5-0.9} and Ridge~3 
areas respectively. The infrared dust map of \cite*{cmotch-WB1:sc01} provides very similar
values.

We show in Fig. \ref{cmotch-WB1_fig:fig3}\ \lnls \ curves for the \object{G21.5-0.9} field
together with the extragalactic ASCA number count relations (\cite{cmotch-WB1:ue99})
extrapolated to faint fluxes. Energy to count factors were computed for the EPIC pn + medium
filter assuming a power law spectrum with photon index 1.9 representative of AGN energy
distributions and a range of N$_{\rm H}$ of a few 10$^{22}$\,cm$^{-2}$. Note that for the same
count rate, a much softer energy distribution such as that emitted by stellar coronae would
imply a factor of $\sim$ 5 fainter X-ray flux in the 0.5-4.5\,keV range. The \lnls \ curves have
not been corrected for full coverage and completeness and are unreliable below
$\sim$ 2 10$^{-14}$\,erg\,cm$^{-2}$\,s$^{-1}$ (0.5-4.5\,keV) and $\sim$ 3
10$^{-14}$\,erg\,cm$^{-2}$\,s$^{-1}$ (4.5-7.5\,keV).  

In spite of possible inaccuracies in the various flux calibrations, it seems that at our flux
completeness level, sources detected in the 0.5-4.5\,keV range are not yet dominated by the
extragalactic population (see Fig. \ref{cmotch-WB1_fig:fig3} left panel). For the N$_{\rm H}$
range derived in the \object{G21.5-0.9} field, the extragalactic component could at best
account for $\sim$ 10\% of the 0.5-4.5\,keV sources at 3 10$^{-14}$ \,erg\,cm$^{-2}$\,s$^{-1}$.
This conclusion is independently supported by the HR2 distribution (see Fig.
\ref{cmotch-WB1_fig:fig2}) since only few sources display a HR2 harder than the minimum value
of $\sim$ 0.7 expected for an AGN seen through N$_{\rm H}$ $\geq$ 5 10$^{22}$\,cm$^{-2}$.
Assuming that some sources with HR2 $\geq$ 0.0 are extragalactic (N$_{\rm H}$ $\geq$ 2
10$^{22}$\,cm$^{-2}$) raises the AGN contribution but not to a point where it can dominate the
source content. 

However, as suggested in Fig. \ref{cmotch-WB1_fig:fig3}, the very faint end of the 
4.5-7.5\,keV hard sources could well be dominated by the AGN background. A modest extragalactic
background of sources in our XMM-Newton fields does not contradict the result recently
reported by \cite*{cmotch-WB1:eb01} based on deep Chandra observations of similar low
latitude directions. The 2-10\,keV energy range in which the extragalactic dominates over
galactic source population is 3 10$^{-15}$ to $\sim$ 3 10$^{-14}$ \,erg\,cm$^{-2}$\,s$^{-1}$
while our estimated completeness level in the same energy range is only a few 10$^{-14}$
\,erg\,cm$^{-2}$\,s$^{-1}$.

\begin{figure*}[ht]
  \begin{center}
    \begin{tabular}{cc}
    \psfig{file=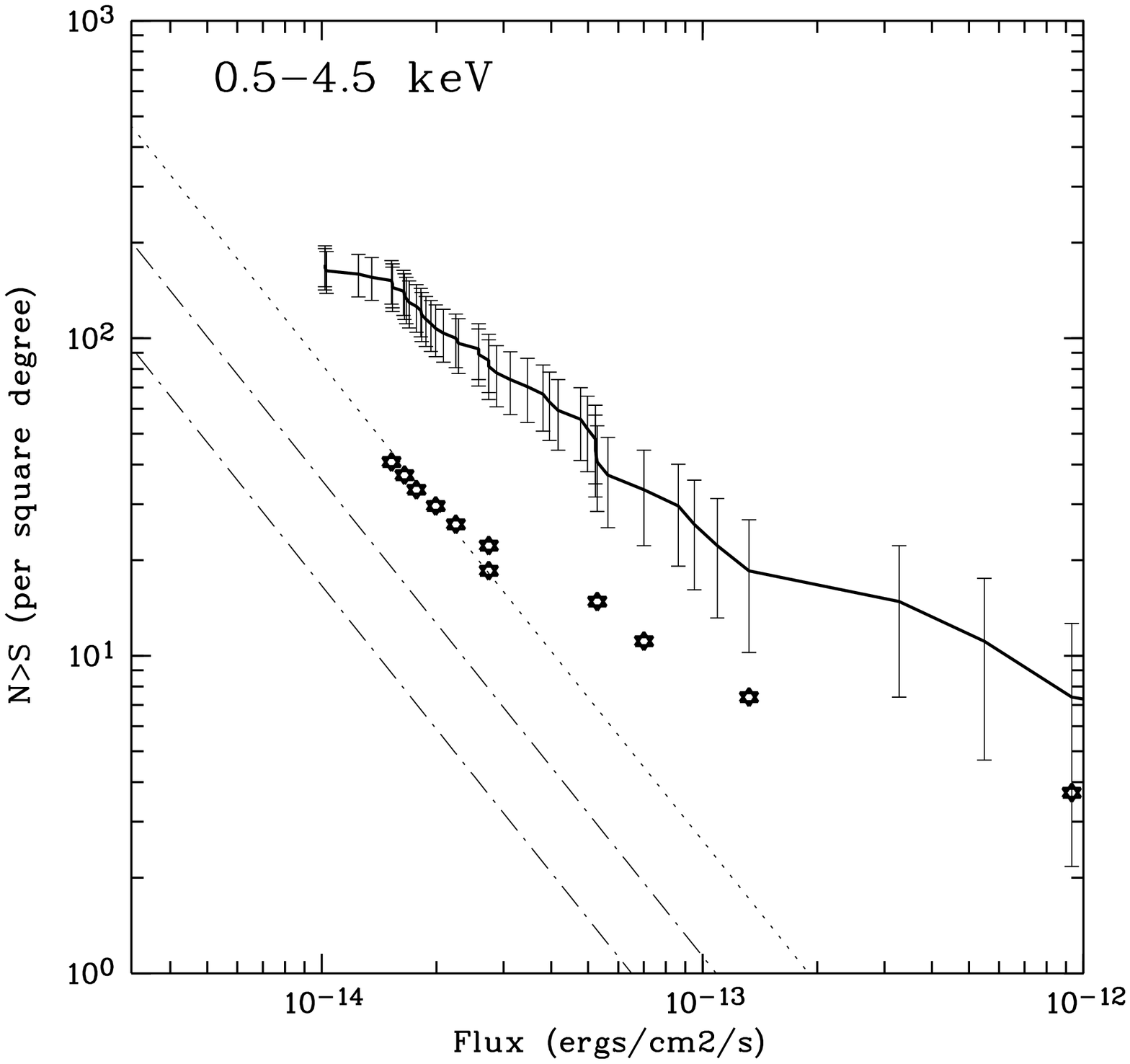, width=8cm, clip=true} &
    \psfig{file=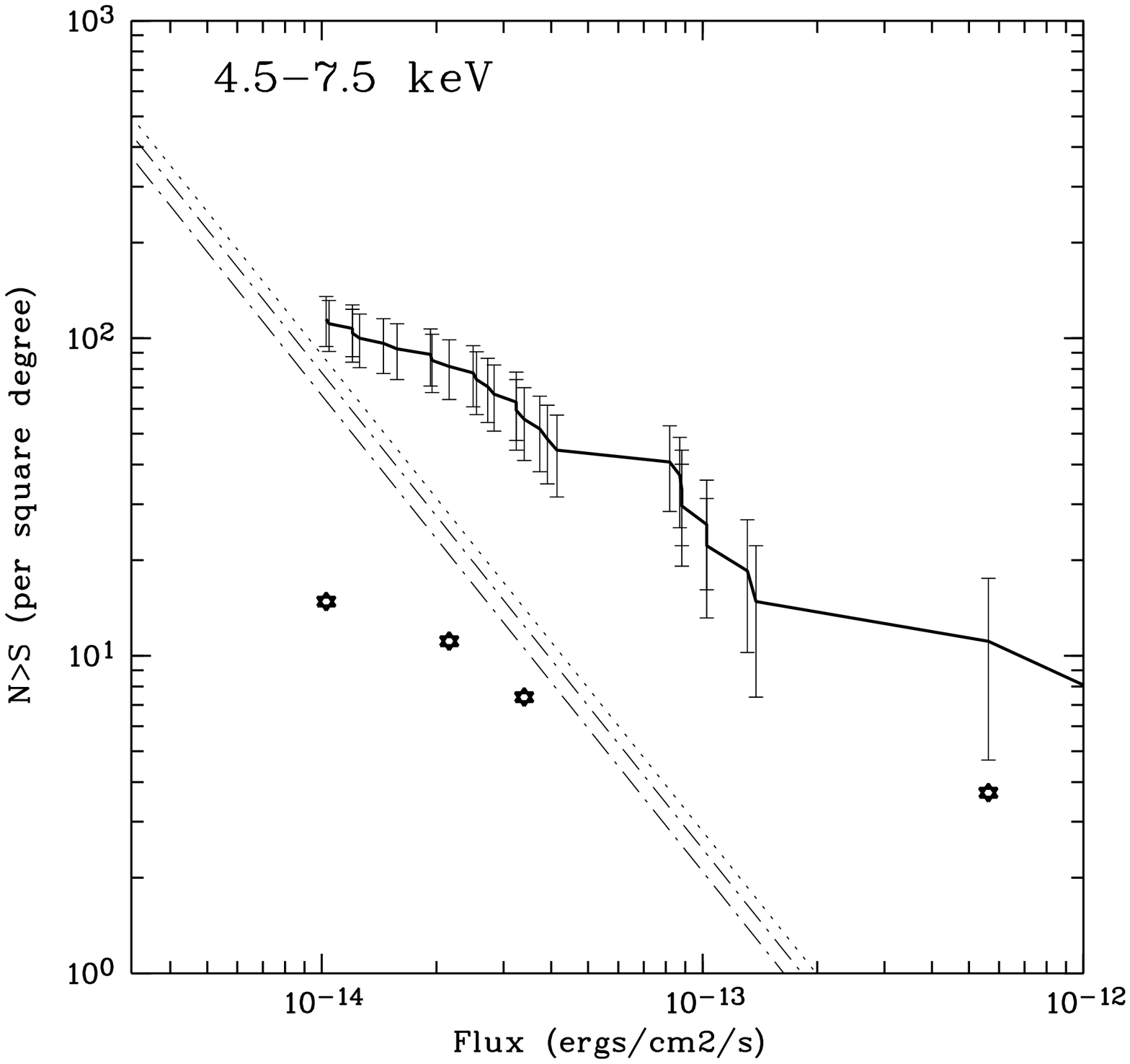, width=8cm, clip=true} \\
    \end{tabular}
  \end{center}
  
\caption{Left: \lnls \ (0.5-4.5\,keV) curve for XMM-Newton sources in the \object{G21.5-0.9}
field (thick line) computed using an energy to count factor typical for background AGN. Stars
represent the number count relation for active coronae. The extragalactic ASCA \lnls\ absorbed
by N$_{\rm H}$ = 5 and 9 10$^{22}$\,cm$^{-2}$, the value suggested by HI and CO maps, are shown
as long dashed lines while that computed assuming N$_{\rm H}$ = 2 10$^{22}$\,cm$^{-2}$,
compatible with the HR2 distribution, is the dotted line. Right: Same figure in the 4.5 to 
7.5\,keV energy range} 

\label{cmotch-WB1_fig:fig3}
\end{figure*}

\section{Conclusions}

In the present state of our work, the nature of the unidentified sources which still
represent the majority of the cases remains unclear. Because of the patchiness of the cold
interstellar medium the density of background AGN may be slightly underestimated but
apparently not to the point where it can dominate source counts. It is fairly possible that a
large number of active coronae have escaped our investigations. We expect some EPIC sources
to be identified with young active F-K stars or RS CVn binaries fainter than the R $\sim$ 17
limit set for chance position coincidence by the still relatively large error circles and
high field stellar densities. Detecting the sometimes subtle signature of their X-ray
activity, such as close binarity, filled H$\alpha$ line or Ca II H\&K emission may be very
consuming in optical observing time. We may have also missed some of the optically faintest
Me stars below R = 19. Part or all of the soft HR2 unidentified sources seen in
Fig.\ref{cmotch-WB1_fig:fig2} could well be optically faint active coronae. 

Cataclysmic variables may also contribute to some extent. They can exhibit a wide range of
X-ray spectra and although their X-ray \lnls \ remains uncertain (\cite{cmotch-WB1:wa99})
they could account for up to $\sim$ 10 of the 70 sources in the field of \object{G21.5-0.9}
while still being optically fainter than R $\sim$ 19. Both Me stars and CVs are strong Balmer
lines emitters and the deep narrow band H$\alpha$ imaging currently carried out will provide
strong constraints on the density of these populations. 

The lower identification rate and stellar fraction in the \object{G21.5-0.9} field compared to that
of Ridge~3 is probably due to a nearby absorbing structure clearly visible on
the DSS-2 images and reflected in the fact that at R = 16 the cumulative density of stars in
Ridge~3 is twice that of \object{G21.5-0.9}. The local cloud in \object{G21.5-0.9} could also account for the
generally harder HR2 distribution in that field.

\end{document}